# Surface Phonons in the Topological Insulators $Bi_2Se_3$ and $Bi_2Te_3$

Ibrahim Boulares,[1] Guangsha Shi,[2] Emmanouil Kioupakis,[2] Petr Lošťák,[3]

Ctirad Uher,[1] and Roberto Merlin[1]

[1]*Department of Physics, University of Michigan, Ann Arbor, Michigan 48109-1040, USA*

[2]*Department of Materials Science and Engineering, University of Michigan, Ann Arbor,*

*Michigan 48109-2136, USA*

[3]*University of Pardubice, Faculty of Chemical Technology, Studentska 573, 53210 Pardubice,*

*Czech Republic*

Raman scattering [K. M. F. Shahil et al., Appl. Phys. Lett. **96**, 153103 (2010), V. Gnezdilov et al., Phys. Rev. B **84**, 195118 (2011) and H. –H. Kung et al., Phys. Rev. B **95**, 245406 (2017)], inelastic helium scattering [X. Zhu et al., Phys. Rev. Lett. **107**, 186102 (2011)] and photoemission experiments [J. A. Sobota et al., Phys. Rev. Lett. **113**, 157401 (2014)] on the topological insulators $Bi_2Se_3$ and $Bi_2Te_3$ show features in the range ~ 50-160 $cm^{-1}$, which have been assigned alternatively to Raman-forbidden, bulk infrared modes arising from symmetry breaking at the surface or to surface phonons, which couple to the topologically protected electronic states. Here, we present temperature- and wavelength- dependent Raman studies showing additional modes we ascribe to surface phonons in both $Bi_2Se_3$ and $Bi_2Te_3$. Our assignment is supported by density functional theory calculations revealing surface phonons at frequencies close to those of the extra peaks in the Raman data. The theoretical results also indicate that these modes are not a consequence of spin-orbit coupling and, thus, that their occurrence is unrelated to the topological properties of these materials.

# I.    INTRODUCTION

Topological insulators (TIs) are a new class of materials that are insulating in the bulk but exhibit metallic surfaces, which arise from strong spin-orbit coupling and particular properties of their band structure. The electronic surface states of TIs consist of gapless bands characterized by a linear (Dirac) dispersion, which are protected from backscattering by time reversal symmetry [1,2]. In recent years, these novel materials have attracted significant interest, not only due to their unique electronic properties, but also because they hold promise for applications in quantum computing [2,3] and spintronic [4], as well as terahertz detection [5], thermoelectric [6] and tunable nonlinear optical devices [7].

$Bi_2Se_3$ and $Bi_2Te_3$ are layered compounds, which have been extensively studied in the past due to their exceptional thermoelectric properties [8]. They were also among the first compounds identified as three-dimensional TIs [9,10,11,12]. Because of its crucial relevance to their surface conductivity properties, the study of electron-phonon coupling [13] and, moreover, the search for vibrations localized at the surface have been the subject of many studies in recent years [14,15,16,17]. In particular, inelastic helium scattering [14,15], surface enhanced Raman scattering [16] and time-resolved photoemission measurements [17] in $Bi_2Se_3$ and $Bi_2Te_3$ show features that were attributed to surface modes as well as strong electron-phonon coupling at the surface. Also, weak features observed in Raman spectra were attributed to surface effects unrelated to the topological surface states [18].

Here, we present experimental results on the temperature- and excitation-wavelength ($\lambda_L-$) dependence of Raman scattering, as well as first-principles phonon calculations for bulk and few-quintuple-layer $Bi_2Se_3$ and $Bi_2Te_3$. Other than the expected, and previously reported Raman-active bulk modes [19,20], we find weak peaks at low temperatures in both compounds, which we ascribe to surface vibrational modes. Density functional theory calculations, which do not consider spin-orbit coupling, support such an assignment in that they reveal a pair of surface-modes, the lower-frequency of which is very close in frequency with the peaks found in the Raman experiments. Arguments are also given suggesting that spin-orbit effects are not important in determining the structural properties and phonon dispersion in these materials.

## II. EXPERIMENTAL AND THEORETICAL METHODS

Raman spectra were obtained in the backscattering geometry with the scattering wavevector along the $\hat{c}$ axis, for temperatures in the range 10-130 K. We used a double grating spectrometer (Dilor XY) and imaged the spectra on a CCD camera (Synapse Horiba). As sources, we employed an argon ion laser, a rhodamine and DCM dye laser and a Ti: sapphire cw laser. The samples were cleaved in air before the measurements and immediately placed under vacuum in a Janis ST-300 cryostat. Data are reported for single crystals of $Bi_2Se_3$ and $Bi_2Te_3$ grown by the Bridgman-Stockbarger technique. We note that low temperature Raman data for single crystal $Bi_2Te_3$ has not been reported prior to this work.

First-principles calculations of bulk and surface phonons of $Bi_2Se_3$ and $Bi_2Te_3$ were carried out using density functional theory [21], with norm-conserving pseudopotentials [22] and a plane-wave cutoff of 50 Ry, as implemented in the Quantum-Espresso code [23]. All calculations use the local-density approximation [24] for the exchange-correlation potential. We sampled the Brillouin zone using a shifted $k$-point grid of 10×10×10 for the bulk materials, and of 8×8×1 grid for the few-layer structures. The bulk and few-layer structures were relaxed with a convergence threshold of $10^{-6}$ Ry/$a_0$ for the forces on the atoms and $10^{-8}$ Ry for the total energy. Phonon frequencies and dynamical matrices were obtained using density-functional perturbation theory [25]. We included the non-analytic term to account for the splitting between the transverse-optical (TO) and longitudinal-optical (LO) modes at the $\Gamma$ point of the Brillouin zone. For bulk materials, we used trigonometric polynomials to interpolate the phonon frequencies to 1000 points along the $\Gamma$–$Z$ direction and subsequently calculated the density of states (DOS) corresponding to the $\Gamma$–$Z$ modes using Gaussian functions and a broadening parameter of 0.2 cm$^{-1}$. Surface phonons of $Bi_2Se_3$ and $Bi_2Te_3$ were studied for 6-quintuple-layer slab structures.

## III. RESULTS AND DISCUSSION

$Bi_2Se_3$ and $Bi_2Te_3$ both crystalize in the space group $R\overline{3}m$, with point group $D_{3d}$. With 5 atoms in the rhombohedral unit cell, these materials possess twelve optical phonon modes at the center of the Brillouin zone. These modes transform as $2E_g + 2A_{1g} + 2E_u + 2A_{1u}$, where $2E_g + 2A_{1g}$ and $2E_u + 2A_{1u}$ are, respectively, Raman and infrared-active representations [19]. We note that the bulk optical phonons of $Bi_2Se_3$ and $Bi_2Te_3$ have been previously measured using Raman [20,26,27,28] and neutron scattering [29], and infrared spectroscopy [19,20,30],

Figure 1 shows Raman spectra of $Bi_2Se_3$ and $Bi_2Te_3$ at various temperatures in the $z(yy)\bar{z}$ ($A_g + E_g$) scattering configuration; note the logarithmic scale. Consistent with previous reports [20,27], the bulk Raman-active modes at 10 K are at 38.5 cm$^{-1}$ ($E_g^1$), 75.5 cm$^{-1}$ ($A_{1g}^1$), 135.8 cm$^{-1}$ ($E_g^2$) and 178.2 cm$^{-1}$ ($A_{1g}^2$) for $Bi_2Se_3$ and 64.1 cm$^{-1}$ ($A_{1g}^1$) 106.1 cm$^{-1}$ ($E_g^2$) and 139.0 cm$^{-1}$ ($A_{1g}^2$) for $Bi_2Te_3$; see Table I. The spectra in the insets were obtained in the $E_g$ (red, top) and $E_g + A_g$ (blue, bottom) configurations. The peak we attribute to a surface phonon is the weak feature which appears between the $E_g^2$ and $A_{1g}^2$ modes at 159 cm$^{-1}$ and 114 cm$^{-1}$ for $Bi_2Se_3$ and $Bi_2Te_3$, respectively. This peak has been previously observed in thin films and few-monolayer samples of both compounds [16,18,31,32,33,34,35,36,37,38,39] and in single crystals of $Bi_2Se_3$ [39,40], but never before in bulk $Bi_2Te_3$. Note that these prior studies detected numerous other weak features in both compounds that were not observed in the present work.

The peak we assign to a surface phonon appears only in the $A_{1g}$ scattering representation, and is strongest at 10 K. In both compounds, the peak intensity decreases with increasing temperature and nearly vanishes above 130 K. Figure 2 shows Raman spectra at 10 K at various excitation wavelengths; the intensity scale is linear. Note the presence of the surface mode at all wavelengths. Similar to the bulk phonons, the Raman cross section for the extra peak depends weakly on $\lambda_L$. The inset in Fig. 2 (a) shows an enlarged view of the $\lambda_L = 780$ nm spectrum. The surface phonon clearly shows an asymmetric lineshape, which we tentatively attribute to Fano-type interference due to coupling to a continuum [41].

As mentioned earlier, features which cannot be attributed to bulk vibrational modes have been previously reported in Raman studies of $Bi_2Se_3$ and $Bi_2Te_3$. Their origin remains controversial. Spectra of few-monolayer $Bi_2Te_3$ show lines at ~ 93 $cm^{-1}$ and 114 $cm^{-1}$ [18,31,32,33,34,35] and at ~ 160 $cm^{-1}$ in $Bi_2Se_3$ [36,37], which were assigned to Raman-forbidden, infrared modes resulting from surface-induced symmetry breaking, while thin-film studies of $Bi_2Te_3$ reveal five additional peaks, one of which, at ~ 93 $cm^{-1}$, was assigned to a surface mode [38]. In bulk, single crystal of $Bi_2Se_3$, additional peaks at 68, 125, 129 and 160 $cm^{-1}$ were observed and also assigned to Raman-forbidden polar modes [40]. Time-resolved photoemission data from single crystal $Bi_2Se_3$ shows an additional mode at 68.4 $cm^{-1}$ [17], which was ascribed to a surface phonon strongly coupled to the surface electronic states. A detailed Raman study of thin films and single crystals of $Bi_2Se_3$ shows four additional modes that were assigned to surface phonons associated with particular bulk branches; their appearance was attributed to out-of-plane lattice distortions which are known to occur at the surface of the crystal. [39].

The observation of extra Raman lines and, in particular, the lines we assign to surface phonons have been prevalently attributed in the literature to surface-induced symmetry breaking, which seemingly allows for scattering of nominally Raman forbidden, infrared-active modes [18,31,32,33,34,35,36,38,40]. For various reasons, we find such an interpretation to be incorrect. The frequencies of TO and LO phonons obtained from fits to infrared reflectivity spectra using a single-oscillator mode [20] are listed in Table I. For $Bi_2Te_3$, we find that the position of the extra Raman line, at 114 cm$^{-1}$, is very close to that of the transverse component of the $A_{1u}^2$ mode. However, assigning the extra peak to such a TO mode would be inconsistent with the facts that (*i*) the width of this phonon is 6 cm$^{-1}$ [20], which is approximately twice that of the Raman peak 3 cm$^{-1}$, and (*ii*) its direction of propagation is perpendicular to the scattering wavevector, which is along [111]. To the best of our knowledge, experimental values of $A_{1u}$ (infrared) phonon frequencies in $Bi_2Se_3$ are not available. Thus, one can only accurately state that the line we ascribe to a surface mode, at 159 cm$^{-1}$, does not match the values for $E_u$ TO or LO modes. Nevertheless, given that the extra modes in $Bi_2Se_3$ and $Bi_2Te_3$ exhibit very similar behavior in regard to selection rules, temperature and $\lambda_L$ dependence, we believe that the possibility that the $Bi_2Se_3$ Raman line at 159 cm$^{-1}$ could be due to a TO or LO $A_{1u}$ -mode is highly unlikely. Finally, we recall that the so-called "forbidden" LO-scattering, which relies on the Fröhlich interaction, is the only known mechanism by which a forbidden infrared mode can become Raman allowed [42], and that such a process displays a strong resonant enhancement. Since the dependence of the intensity of the extra peaks on $\lambda_L$ is rather weak (see Fig. 2), and consistent with the absence of a frequency match with LO phonons in $Bi_2Te_3$, it follows that the additional lines in either compound cannot be ascribed to forbidden LO scattering.

The results of our first-principles calculations support our contention that the additional peaks we observe in $Bi_2Se_3$ and $Bi_2Te_3$ are due to surface phonons. Fig. 3 shows the calculated frequencies and atomic displacements of surface phonons in both compounds from slab calculations, along with the $\Gamma$–Z bulk-phonon projected DOS. We see two surface modes (160.4 $cm^{-1}$ and 183.2 $cm^{-1}$ for $Bi_2Se_3$, and 117.5 $cm^{-1}$ and 146.8 $cm^{-1}$ for $Bi_2Te_3$). The calculated frequency of the low-frequency mode is in good agreement with experimental measurements in both compounds. The calculations show that these modes are strongly localized on the topmost quintuple layer of the slab structure, and that the calculated displacements, along the $\hat{c}$ axis, are in very good agreement with our experimental observation that the symmetry of these modes is $A$. The fact that their frequencies fall in the gap regions of the corresponding bulk projected DOS and, thus, that they do not mix with bulk modes folded along the $\Gamma$–Z direction indicates that these modes are truly surface modes, as opposed to surface resonances. Our experimental data shows no clear evidence of the high frequency mode, which is expected to be in close proximity to and could be hidden underneath the strong, bulk $A_{1g}^2$ phonon.

Spin-orbit interaction was not considered in the evaluation of the vibrational frequencies. Although spin-orbit coupling strongly affects the band structure of semiconductors with heavy elements such as Bi [43], and gives rise to the topologically insulating behavior [44], it is not important for the evaluation of structural properties or vibrational frequencies. The effect of spin-orbit coupling is secondary to the choice of the exchange-correlation function in affecting the accuracy of the calculated frequencies. For example, the zone-center phonon frequencies of $Bi_2Se_3$ and $Bi_2Te_3$ have been calculated [45] using the local density approximation (LDA) [46] and the Perdew-Burke-Exchange (PBE) [47] functionals. For both functionals, results were obtained both with and without spin-orbit coupling effects included in the calculation. The results show that the root mean square difference of the calculated frequencies determined with LDA upon inclusion of spin-orbit coupling (5.14 cm$^{-1}$) is smaller than the difference between the results obtained using the two different functionals (5.92 cm$^{-1}$ when ignoring and 7.53 cm$^{-1}$ when including spin-orbit coupling effects). In Table I, we list the bulk phonon frequencies from our theoretical and experimental work, as well as previously reported LDA calculations [45], which include spin-orbit coupling and infrared measurements [20] of phonon frequencies. The discrepancies between theory and experiment for the infrared modes may be attributed to the approximations involved in density functional theory (i.e.., the choice of the exchange-correlation functional and pseudopotentials) or due to higher-order effects such as phonon anharmonicity. We note that previous density functional theory calculations also show larger discrepancies with experiment for infrared modes [45].

## IV.    CONCLUSION

In summary, we performed temperature- and excitation wavelength- dependent Raman scattering measurements, as well as density functional theory calculations in the topological insulators $Bi_2Se_3$ and $Bi_2Te_3$. In both compounds, and in addition to the features corresponding to the bulk Raman-allowed modes, our Raman measurements reveal a weak peak at low temperatures, which we assign to surface phonons. First-principles calculations, which do not include spin-orbit coupling, reveal two phonons localized at the topmost layers, one of which coincides in frequency with the extra peak and has a symmetry that is consistent with experimental results in both compounds.

## ACKNOWLEDGEMENTS

The density functional theory calculations were supported by the National Science Foundation CAREER award through Grant N. DMR-1254314. This research used resources of the Na-tional Energy Research Scientific Computing Center, a DOE Office of Science User Facility supported by the office of Science of the U.S. Department of Energy under Contract No. DE-AC02-05CH11231

# Table I

| Symmetry | Bi$_2$Se$_3$ | | | Bi$_2$Te$_3$ | | |
|---|---|---|---|---|---|---|
| | Experiment | Theory | | Experiment | Theory | |
| | | This work | LDA+SOC [45] | | This work | LDA+SOC [45] |
| $E_g^1$ | 39[+] | 45 | 43 | -- | 43 | 42 |
| $A_{1g}^1$ | 75[+] | 75 | 75 | 64[+] | 65 | 63 |
| $E_g^2$ | 136[+] | 143 | 138 | 106[+] | 113 | 105 |
| $A_{1g}^2$ | 178[+] | 183 | 175 | 139[+] | 142 | 132 |
| $E_u^1$ (TO) | 61* | 90 | 82 | 48* | 71 | 64 |
| $E_u^1$ (LO) | 117** | 133 | -- | 86** | 99 | -- |
| $A_{1u}^1$ (TO) | -- | 145 | 137 | 88* | 105 | 97 |
| $A_{1u}^1$ (LO) | -- | 158 | -- | 98** | 114 | -- |
| $E_u^2$ (TO) | 134* | 136 | 131 | 98* | 101 | 95 |
| $E_u^2$ (LO) | 138** | 158 | -- | 100** | 116 | -- |
| $A_{1u}^2$ (TO) | -- | 166 | 163 | 114* | 128 | 121 |
| $A_{1u}^2$ (LO) | -- | 173 | -- | 124** | 134 | -- |

**Table I.** Experimental and theoretical values of Raman and infrared bulk phonon frequencies for Bi$_2$Se$_3$ and Bi$_2$Te$_3$ in units of cm$^{-1}$. Theoretical values including spin-orbit coupling (SOC) are from [45]. Experimental Raman ([+]) and infrared (*) [20] values were measured at 10 K and 15 K, respectively. Frequencies of the LO components of infrared-active phonons (**) were extracted from infrared reflectivity data using a single-oscillator model. Values for Bi$_2$Te$_3$ are in excellent agreement with inelastic neutron scattering results [29].

# Figure Captions

**FIG. 1.** Raman spectra of (a) $Bi_2Se_3$, 488nm excitation, and (b) $Bi_2Te_3$, 780nm excitation, at various temperatures. Selection rules are shown in the insets. The intensity scale is logarithmic for all the traces. Arrows denote the surface mode.

**FIG. 2.** (a) Raman spectra of $Bi_2Se_3$ at various excitation wavelengths. The spectrum in the inset shows Fano-type interference. (b) Data for $Bi_2Te_3$. The intensity scale is linear in all cases.

**FIG. 3.** Calculated frequencies and corresponding atomic displacements for the two surface modes material from slab calculations: (a) 160.4 $cm^{-1}$ and 183.2 $cm^{-1}$ for $Bi_2Se_3$, and (b) 117.5 $cm^{-1}$ and 146.8 $cm^{-1}$ for $Bi_2Te_3$. The atomic displacements of the surface modes are localized primarily in the topmost quintuple layer. Calculated bulk phonon densities of states (DOS), projected along the $\Gamma$–$Z$ direction, are also shown.

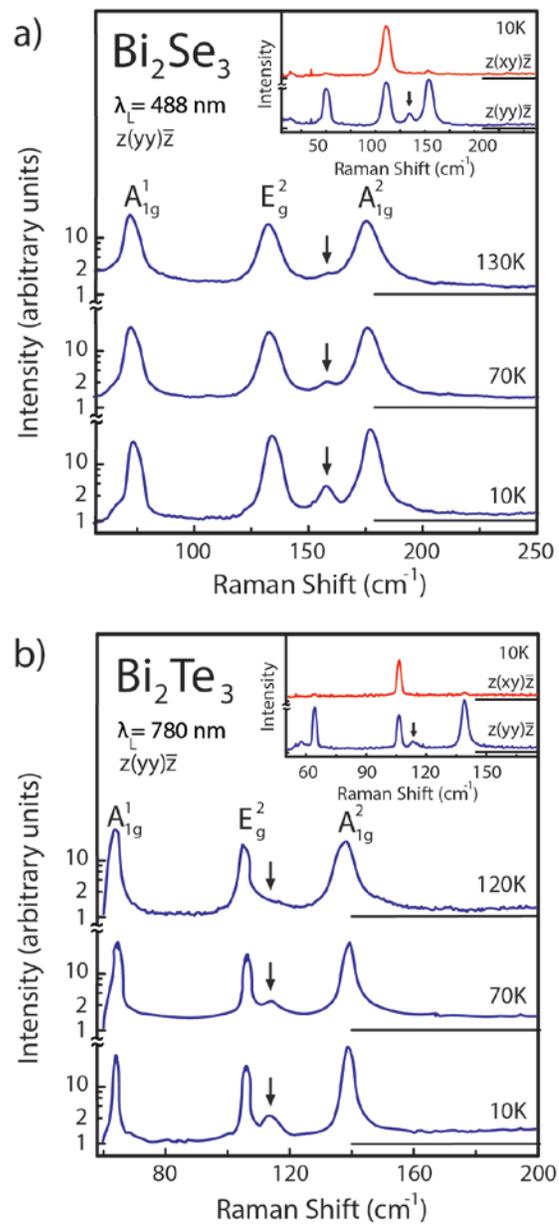

FIGURE 1

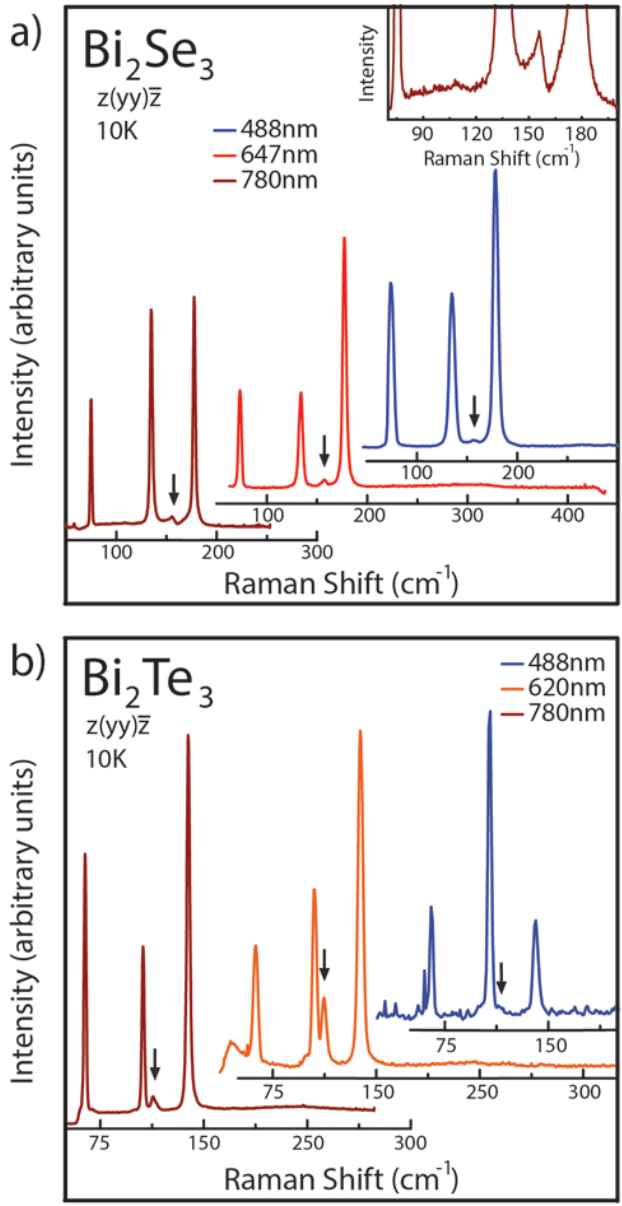

FIGURE 2

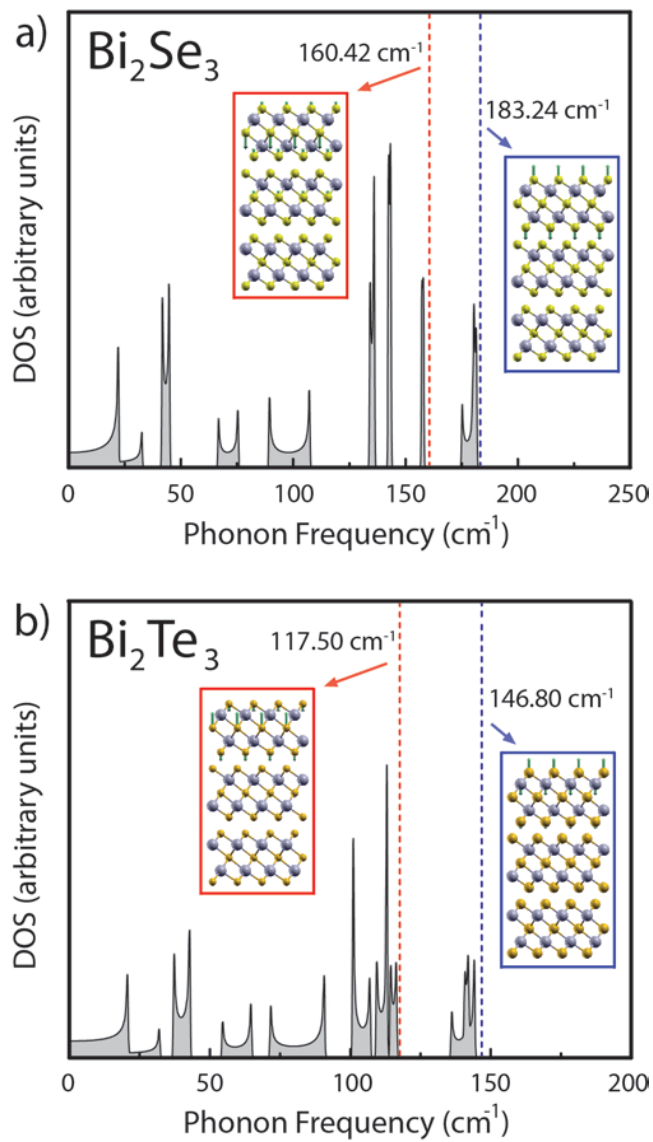

FIGURE 3